\documentclass[10pt,twocolumn,english,final,5p,times]{revtex4}
\usepackage[T1]{fontenc}
\usepackage[latin9]{inputenc}
\usepackage{color}
\usepackage{amsmath}
\usepackage{amssymb}
\usepackage{graphicx}

\makeatletter
\@ifundefined{textcolor}{}
{%
 \definecolor{BLACK}{gray}{0}
 \definecolor{WHITE}{gray}{1}
 \definecolor{RED}{rgb}{1,0,0}
 \definecolor{GREEN}{rgb}{0,1,0}
 \definecolor{BLUE}{rgb}{0,0,1}
 \definecolor{CYAN}{cmyk}{1,0,0,0}
 \definecolor{MAGENTA}{cmyk}{0,1,0,0}
 \definecolor{YELLOW}{cmyk}{0,0,1,0}
}

\usepackage{babel}

\makeatother

\usepackage{babel}
\begin{document}

\title{Thermal entanglement and sharp specific-heat peak in an exactly solved
spin-1/2 Ising-Heisenberg ladder with alternating Ising and Heisenberg
inter-leg couplings}

\author{Onofre Rojas$^{1}$, J. Stre\v{c}ka$^{2}$ and S. M. de Souza$^{1}$}

\address{$^{1}$Departamento de Física, Universidade Federal de Lavras, 37200-000,
Lavras-MG, Brazil}

\address{$^{2}$Institute of Physics, Faculty of Science, P. J. Šafárik University,
Park Angelinum 9, 040 01 Košice, Slovakia}
\begin{abstract}
The spin-1/2 Ising-Heisenberg two-leg ladder accounting for alternating
Ising and Heisenberg inter-leg couplings in addition to the Ising
intra-leg coupling is rigorously mapped onto to a mixed spin-(3/2,1/2)
Ising-Heisenberg diamond chain with the nodal Ising spins $S=3/2$
and the interstitial spin-1/2 Heisenberg dimers. The latter effective
model with higher-order interactions between the nodal and interstitial
spins is subsequently exactly solved within the transfer-matrix method.
The model under investigation exhibits five different ground states:
ferromagnetic, antiferromagnetic, superantiferromagnetic and two types
of frustrated ground states with a non-zero residual entropy. A detailed
study of thermodynamic properties reveals an anomalous specific-heat
peak at low enough temperatures, which is strongly reminiscent because
of its extraordinary height and sharpness to an anomaly accompanying
a phase transition. It is convincingly evidenced, however, that the
anomalous peak in the specific heat is finite and it comes from vigorous
thermal excitations from a two-fold degenerate ground state towards
a macroscopically degenerate excited state. Thermal entanglement between
the nearest-neighbor Heisenberg spins is also comprehensively explored
by taking advantage of the concurrence. The threshold temperature
delimiting a boundary between the entangled and disentangled parameter
space may show presence of a peculiar temperature reentrance. 
\end{abstract}

\keywords{Ising-Heisenberg model, spin ladder, specific heat, entanglement}

\maketitle

\section{Introduction}

In condensed matter physics, one of the most investigated subjects
is the correlation between parts of composite systems \cite{fulde,fazekas}.
In this sense, it is quite relevant to study the quantum part of these
correlations, the so-called entanglement. Quantum entanglement is
a fascinating feature of the quantum theory due to its nonlocal property
\cite{bell}. Therefore, many researchers have focused in recent years
their attention to quantum entanglement as a potential resource for
quantum computing and quantum information processing \cite{nielsen,amicorev}.

From the practical of view, there exist several real magnetic materials
with obvious quantum manifestations as provided for instance by experimental
representatives of the spin-1/2 quantum Heisenberg ladder \cite{batchelor}.
The most widespread families of the spin-1/2 Heisenberg ladder materials
are cuprates Cu$_{2}$(C$_{5}$H$_{12}$N)$_{2}$Cl$_{4}$ \cite{chiari},
SrCu$_{2}$O$_{3}$ \cite{Hiroi}, (C$_{5}$H$_{12}$N)$_{2}$CuBr$_{4}$
\cite{willett}, and vanadates M$^{2+}$V$_{2}$O$_{5}$ \cite{Onoda},
(VO)$_{2}$P$_{2}$O$_{7}$ \cite{barnes}, which involve Cu$^{2+}$
and V$^{4+}$ magnetic ions as the spin-1/2 carriers. Recently, another
experimental realization of the spin-1/2 Heisenberg two-leg ladder
Cu(Qnx)(Cl$_{1-x}$Br$_{x}$)$_{2}$, where Qnx stands for quinoxaline
(C$_{8}$H$_{6}$N$_{2}$), has opened up a new opportunity to continuously
tune the inter-leg to intra-leg coupling ratio albeit in a relatively
narrow range \cite{simutis}.

Motivated by these experiments, a lot of interest has been devoted
to theoretical investigation of the spin-1/2 Heisenberg ladder models
\cite{batchelor}. A large number of studies aimed at several variants
of the Heisenberg spin ladder have addressed the ground-state properties
\cite{amiri,jaan-ladder}. Besides, the existence of a magnetization
plateau in the spin-1/2 Heisenberg ladder with alternating inter-leg
exchange interactions was investigated by Paparidze and Pogosyan \cite{japaridze}.
There exist even a few generalized version of the $N$-leg spin-$S$
Heisenberg ladders \cite{Ramos}, which were investigated using the
density-matrix renormalization group method.

Frustrated spin ladders accounting for the crossing (next-nearest-neighbor)
interaction were also intensively studied, some recent rigorous results
for a ground-state phase diagram of the spin-1/2 Ising-Heisenberg
ladder of this type can be found in Refs. \cite{strecka-ladders,jozef-ladder-14}.
Exact ground states were also found for a frustrated spin-1/2 Ising-Heisenberg
ladder with the Heisenberg inter-leg coupling, the Ising intra-leg
and crossing couplings. This model is in a certain limit equivalent
to the spin-1/2 Ising-Heisenberg tetrahedral chain, which was also
widely explored \cite{strecka-Vk-Ly,strecka-lyra}.

The spin-1/2 Ising-Heisenberg models being composed of the Ising (classical)
and Heisenberg (quantum) spins \cite{vanden10,sahoo12,bellucci} drew
a special attention also from the experimental side as exemplified
by numerous studies of the Ising-Heisenberg spin chains \cite{vanden10,sahoo12,bellucci,hagiwara-strecka,valverde08,ohanyan09,antonosyan09,canova09,rojas-ohanyan,ana-roj,strecka-jasc05}.
In addition, the Ising-Heisenberg chains may display many intriguing
and unexpected quantum properties \cite{valverde08,ohanyan09,antonosyan09,canova09,rojas-ohanyan,spra}
such as thermal entanglement, intermediate plateaux in low-temperature
magnetization curves \cite{antonosyan09,canova09,rojas-ohanyan} or
non-rational magnetization at zero temperature \cite{bellucci,Ohanyan-prb15}.

In the present work, we will examine the spin-1/2 Ising-Heisenberg
ladder with alternating Ising and Heisenberg inter-leg couplings.
The organization of this paper is as follows. In Sec. 2 we will briefly
describe the model under investigation and its rigorous mapping equivalence
with the mixed-(3/2,1/2) Ising-Heisenberg diamond chain. We will also
establish in Sec. 2 the relevant ground-state phase diagram. Sec.
3 deals with thermodynamics of the investigated model, whereas the
particular attention is paid to a detailed study of temperature dependences
of the specific heat and entropy. Sec. 4 is dedicated to the thermal
entanglement between the nearest-neighbor Heisenberg spins. Finally,
our conclusions are drawn in Sec. 5.

\section{Ising-Heisenberg ladder with alternating inter-leg interactions}

\begin{figure}
\centering\includegraphics[scale=0.35]{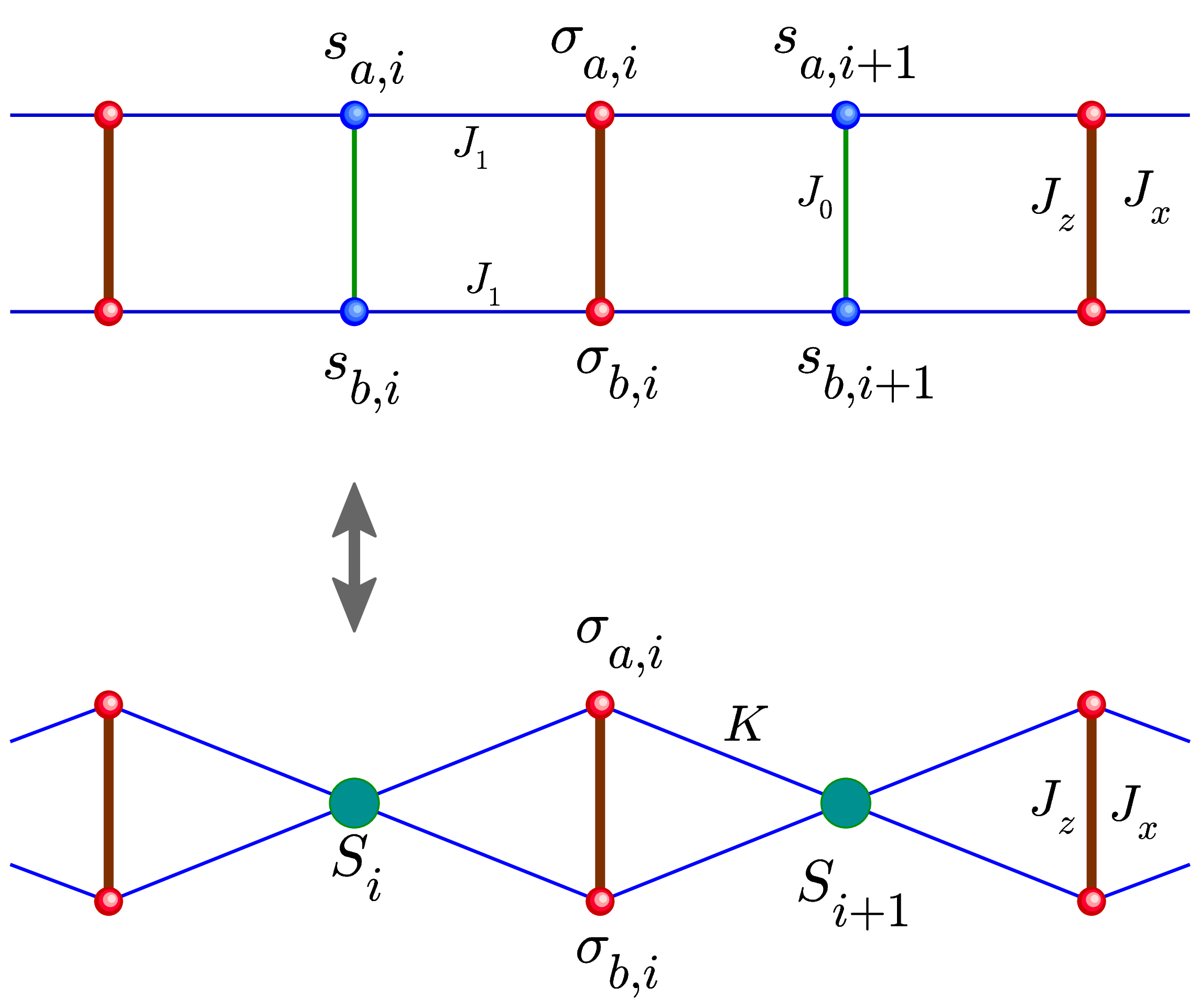}\caption{\label{fig:ladder} (Top) A schematic representation of the spin-1/2
Ising-Heisenberg ladder with alternating Ising and Heisenberg inter-leg
interactions. Thick vertical lines correspond to the Heisenberg coupling
($J_{x},J_{z}$), while thin vertical and horizontal lines correspond
to the Ising interactions $J_{0}$ and $J_{1}$; (Bottom) The equivalent
mixed spin-(3/2,1/2) Ising-Heisenberg diamond chain with the nodal
spin-3/2 Ising spins and the interstitial spin-1/2 Heisenberg dimers.}
\end{figure}

Let us consider the spin-1/2 Ising-Heisenberg ladder accounting for
alternating Ising and Heisenberg inter-leg couplings in addition to
the Ising intra-leg coupling, which is schematically depicted in figure
\ref{fig:ladder}. The Hamiltonian of the aforedescribed spin-1/2
Ising-Heisenberg ladder is given by 
\begin{equation}
\mathcal{H}=\sum_{i=1}^{N}\left(H_{i}^{XXZ}+H_{i,i+1}^{I}+H_{i,i+1}^{IH}\right),\label{eq:Ham-orig}
\end{equation}
where 
\begin{alignat}{1}
H_{i}^{XXZ}= & -J_{x}(\sigma_{a,i}^{x}\sigma_{b,i}^{x}+\sigma_{a,i}^{y}\sigma_{b,i}^{y})-J_{z}\sigma_{a,i}^{z}\sigma_{b,i}^{z},\label{eq:Ht-XXZ}\\
H_{i,i+1}^{I}= & -\frac{J_{0}}{2}(s_{a,i}s_{b,i}+s_{a,i+1}s_{b,i+1}),\label{eq:Ht-I}\\
H_{i,i+1}^{IH}= & -J_{1}(s_{a,i}+s_{a,i+1})\sigma_{a,i}^{z}-J_{1}(s_{b,i}+s_{b,i+1})\sigma_{b,i}^{z}.\label{eq:Ht-HI}
\end{alignat}
In above, $\sigma_{\gamma,i}^{\alpha}$ denotes spatial components
of the spin-1/2 operator ($\alpha=\{x,y,z\}$) at site $i$, and $\gamma=a$
or $b$ (see figure \ref{fig:ladder}). The Ising inter-leg coupling
is denoted by $J_{0}$, the Ising intra-leg coupling is denoted by
$J_{1}$, while the anisotropic XXZ Heisenberg inter-leg coupling
has two spatial components $J_{x}$ and $J_{z}$ in the $xy$-plane
and along $z$-axis, respectively.

To proceed further with the calculation, let us proof a rigorous mapping
equivalence between the spin-1/2 Ising-Heisenberg ladder defined through
the total Hamiltonian (\ref{eq:Ham-orig}) and the mixed spin-(3/2,1/2)
Ising-Heisenberg diamond chain with the nodal spin-3/2 Ising spins
and the interstitial spin-1/2 Heisenberg dimers as schematically illustrated
in figure \ref{fig:ladder}(bottom). The exact mapping relationship
between both models can be proven by the use of the following spin
identities \cite{nikolay} 
\begin{alignat}{1}
s_{a}(S)= & \frac{13}{12}S-\frac{S}{3}^{3},\\
s_{b}(S)= & \frac{7}{6}S-\frac{2S}{3}^{3},
\end{alignat}
which establish the exact mapping correspondence between the old spin-1/2
Ising variables $s_{a},s_{b}$ and the novel spin-3/2 Ising variable
$S$ 
\begin{alignat}{3}
s_{a}= & \frac{1}{2}, & s_{b} & =-\frac{1}{2} & \qquad\Longleftrightarrow\qquad & S=\frac{3}{2},\\
s_{a}= & \frac{1}{2}, & s_{b} & =\frac{1}{2} & \qquad\Longleftrightarrow\qquad & S=\frac{1}{2},\\
s_{a}= & -\frac{1}{2},\,\,\, & s_{b} & =-\frac{1}{2} & \qquad\Longleftrightarrow\qquad & S=-\frac{1}{2},\\
s_{a}= & -\frac{1}{2}, & s_{b} & =\frac{1}{2} & \qquad\Longleftrightarrow\qquad & S=-\frac{3}{2}.
\end{alignat}
Consequently, the Hamiltonian parts \eqref{eq:Ht-I} and \eqref{eq:Ht-HI}
depending on the old spin-1/2 variables $s_{a},s_{b}$ can be rewritten
in terms of the novel spin-3/2 Ising variables 
\begin{alignat}{1}
H_{i,i+1}^{I}= & J_{0}(\frac{S_{i}^{2}+S_{i+1}^{2}}{8}-\frac{5}{16}),\label{eq:ti}\\
H_{i,i+1}^{IH}= & J_{1}\left(\sigma_{a,i}^{z}+2\sigma_{b,i}^{z}\right)\frac{S_{i}^{3}+S_{i+1}^{3}}{3}\nonumber \\
 & -J_{1}\left(13\sigma_{a,i}^{z}+14\sigma_{b,i}^{z}\right)\frac{S_{i}+S_{i+1}}{12}.\label{eq:tih}
\end{alignat}
In this way, one establishes a rigorous mapping equivalence between
the spin-1/2 Ising-Heisenberg ladder defined by the Hamiltonians (\ref{eq:Ht-XXZ}),
(\ref{eq:Ht-I}), (\ref{eq:Ht-HI}) and, respectively, the mixed spin-(3/2,1/2)
Ising-Heisenberg diamond chain with the nodal Ising spins $S=3/2$
and the interstitial spin-1/2 Heisenberg dimers defined by the effective
Hamiltonians (\ref{eq:Ht-XXZ}), (\ref{eq:ti}), (\ref{eq:tih}).
More importantly, it can be understood from the Hamiltonian (\ref{eq:ti})
that the Ising inter-leg coupling $J_{0}$ gives rise to a uniaxial
single-ion anisotropy acting on the effective spin-3/2 Ising variables,
while the Ising intra-leg coupling $J_{1}$ produces unusual bilinear
and higher-order (quartic) interactions between the Heisenberg and
Ising spins.

\subsection{The ground-state phase diagram}

The spin-1/2 Ising-Heisenberg ladder given by the Hamiltonian \eqref{eq:Ham-orig}
exhibits in a zero magnetic field five different ground states. Two
ground states are classical two-fold degenerate ferromagnetic (FM)
and superantiferromagnetic (SAF) phases given by the eigenvectors
\begin{align}
|FM\rangle= & \Biggl\{\begin{array}{c}
{\displaystyle \prod_{i=1}^{N}|\begin{smallmatrix}+\\
+
\end{smallmatrix}\rangle_{\sigma_{i}}\otimes|\begin{smallmatrix}+\\
+
\end{smallmatrix}\rangle_{s_{i}}}\\
{\displaystyle \prod_{i=1}^{N}|\begin{smallmatrix}-\\
-
\end{smallmatrix}\rangle_{\sigma_{i}}\otimes|\begin{smallmatrix}-\\
-
\end{smallmatrix}\rangle_{s_{i}}}
\end{array}\Biggr.,\\
|SAF\rangle= & \Biggl\{\begin{array}{c}
{\displaystyle \prod_{i=1}^{N}|\begin{smallmatrix}+\\
+
\end{smallmatrix}\rangle_{\sigma_{i}}\otimes|\begin{smallmatrix}-\\
-
\end{smallmatrix}\rangle_{s_{i}}}\\
{\displaystyle \prod_{i=1}^{N}|\begin{smallmatrix}-\\
-
\end{smallmatrix}\rangle_{\sigma_{i}}\otimes|\begin{smallmatrix}+\\
+
\end{smallmatrix}\rangle_{s_{i}}}
\end{array}\Biggr..
\end{align}
To simplify the notation, the former state vector with the subscript
$\sigma_{i}$ corresponds to the $i$th Heisenberg dimer $\sigma_{a,i},\sigma_{b,i}$,
while the latter state vector with the subscript $s_{i}$ corresponds
to the $i$th Ising dimer $s_{a,i},s_{b,i}$. The relevant ground-state
energies per unit cell are given by 
\begin{align}
E_{FM}= & -J_{1}-\frac{1}{4}J_{z}-\frac{1}{4}J_{0},\\
E_{SAF}= & J_{1}-\frac{1}{4}J_{z}-\frac{1}{4}J_{0}.
\end{align}
In addition, there also exist two highly degenerate frustrated ground
states, namely, the quantum frustrated phase FRU1 and the classical
frustrated phase FRU2 given by the eigenvectors 
\begin{align}
|FRU1\rangle= & \prod_{i=1}^{N}|\tau\rangle_{\sigma_{i}}\otimes|\begin{smallmatrix}b\\
b
\end{smallmatrix}\rangle_{s_{i}},\label{eq:SS}\\
|FRU2\rangle= & \prod_{i=1}^{N}|\begin{smallmatrix}a\\
a
\end{smallmatrix}\rangle_{\sigma_{i}}\otimes|\begin{smallmatrix}b\\
-b
\end{smallmatrix}\rangle_{s_{i}}.\label{eq:TT}
\end{align}
Here, the symbols $a$ and $b$ can take any of two possible values
$\pm$ and the symbol $\tau$ refers to 
\begin{alignat}{1}
|\tau\rangle_{\sigma_{i}}= & \frac{1}{\sqrt{2}}\left(|\begin{smallmatrix}+\\
-
\end{smallmatrix}\rangle_{\sigma_{i}}+\mbox{sign}(J_{x})|\begin{smallmatrix}-\\
+
\end{smallmatrix}\rangle_{\sigma_{i}}\right).\label{tau}
\end{alignat}
The corresponding ground-state energies of the FRU1 and FRU2 phases
are given by 
\begin{align}
E_{FRU1}= & -\frac{1}{2}|J_{x}|+\frac{1}{4}J_{z}-\frac{1}{4}J_{0},\\
E_{FRU2}= & -\frac{1}{4}J_{z}-\frac{1}{4}|J_{0}|.
\end{align}
Finally, there exist the peculiar two-fold degenerate quantum antiferromagnetic
(AFM) ground state given by the eigenvector 
\begin{align}
|AFM\rangle= & \Biggl\{\begin{array}{c}
{\displaystyle \prod_{i=1}^{N}|\eta_{+}\rangle_{\sigma_{i}}\otimes|\begin{smallmatrix}+\\
-
\end{smallmatrix}\rangle_{s_{i}}}\\
{\displaystyle \prod_{i=1}^{N}|\eta_{-}\rangle_{\sigma_{i}}\otimes|\begin{smallmatrix}-\\
+
\end{smallmatrix}\rangle_{s_{i}}}
\end{array}\Biggr.,
\end{align}
where 
\begin{align}
|\eta_{+}\rangle_{\sigma_{i}}= & \tfrac{1}{\sqrt{1+c^{2}}}\left(-|\begin{smallmatrix}-\\
+
\end{smallmatrix}\rangle_{\sigma_{i}}+c|\begin{smallmatrix}+\\
-
\end{smallmatrix}\rangle_{\sigma_{i}}\right),\nonumber \\
|\eta_{-}\rangle_{\sigma_{i}}= & \tfrac{1}{\sqrt{1+c^{2}}}\left(|\begin{smallmatrix}+\\
-
\end{smallmatrix}\rangle_{\sigma_{i}}+c|\begin{smallmatrix}-\\
+
\end{smallmatrix}\rangle_{\sigma_{i}}\right),\label{eta}
\end{align}
with 
\begin{equation}
c=\frac{2J_{1}+\sqrt{4J_{1}^{2}+J_{x}^{2}}}{J_{x}}.
\end{equation}
The respective ground-state energy per unit cell of the AFM phase
is given by 
\begin{align}
E_{AFM}= & \frac{1}{4}J_{z}+\frac{1}{4}J_{0}-\frac{1}{2}\sqrt{4J_{1}^{2}+J_{x}^{2}}.
\end{align}

\begin{figure}
\begin{centering}
\includegraphics[scale=0.22]{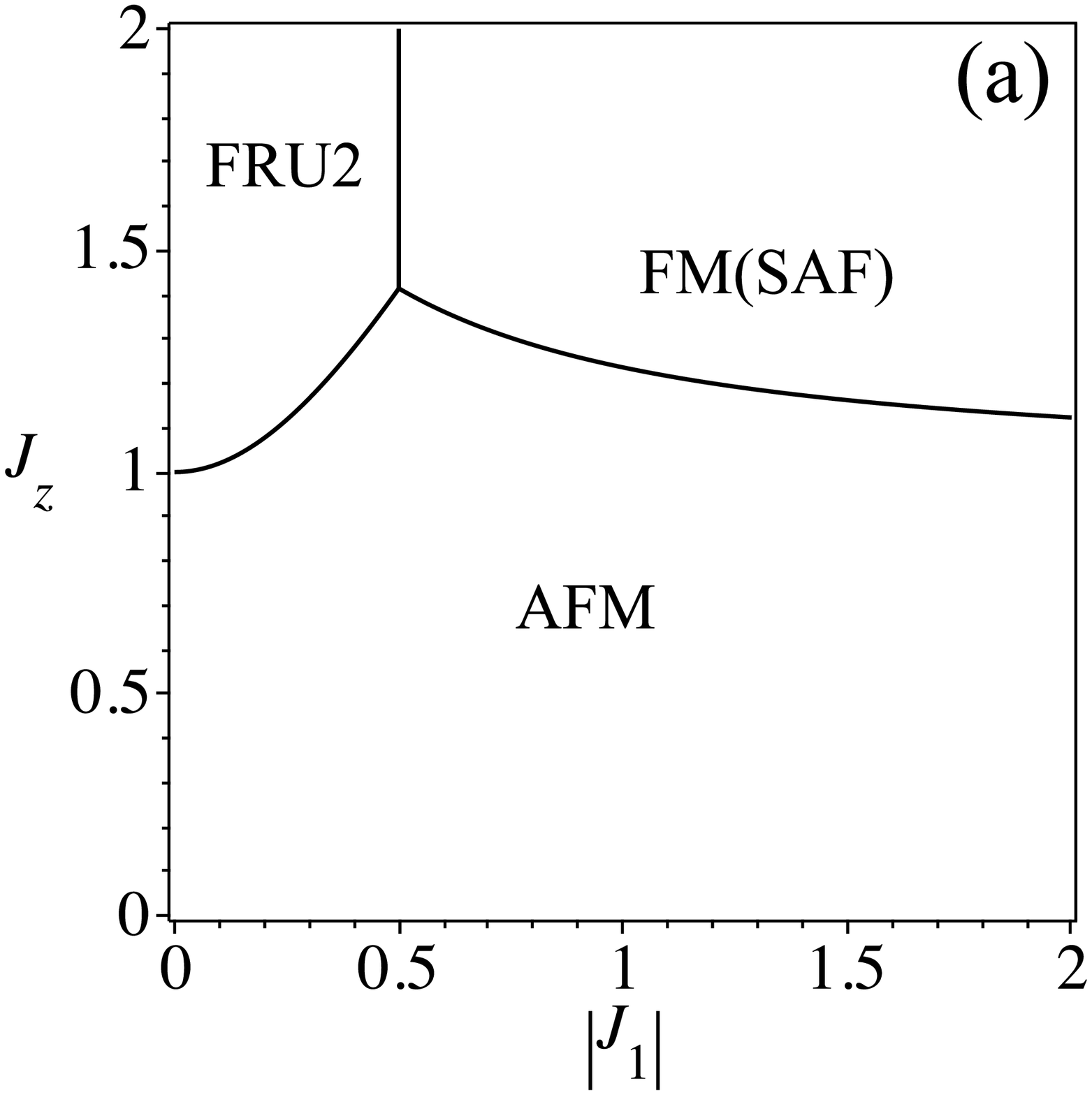}\includegraphics[scale=0.22]{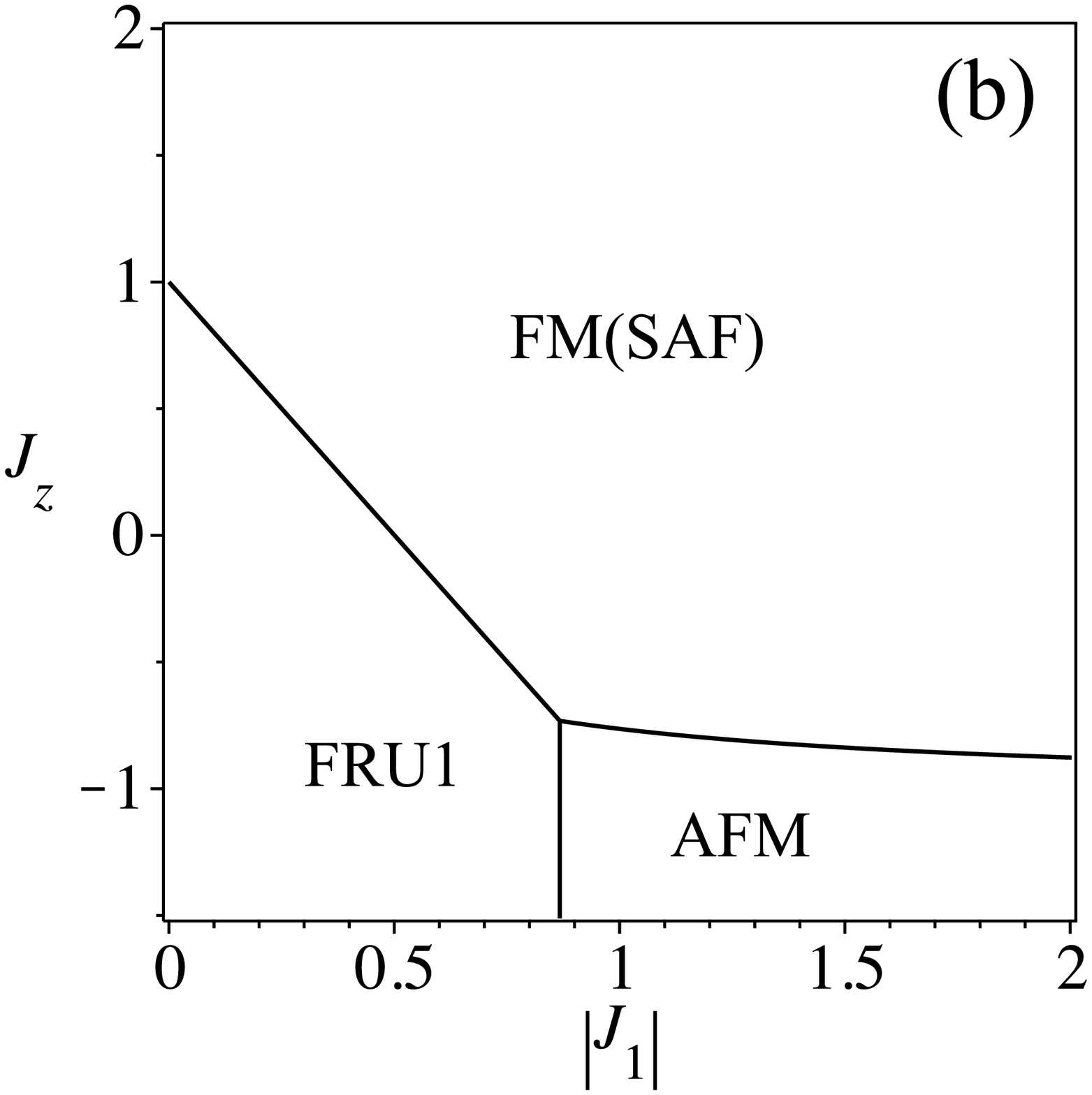} 
\par\end{centering}

\caption{\label{fig:The-phase-diagram}The ground-state phase diagram in the
$|J_{1}|$-$J_{z}$ plane for the fixed values of the coupling constants:
(a) $J_{x}=1$, $J_{0}=-1$; (b) $J_{x}=1$, $J_{0}=1$. }
\end{figure}

In figure \ref{fig:The-phase-diagram} we illustrate the ground-state
phase diagrams in the $|J_{1}|$-$J_{z}$ plane by considering the
fixed values of the coupling constants: (a) $J_{x}=1$, $J_{0}=-1$.
While the change in sign of the transverse component of the Heisenberg
interaction $J_{x}$ is merely responsible for a change of the symmetry
of the eigenvectors (\ref{tau}) and (\ref{eta}), the change in character
of the Ising inter-leg coupling $J_{0}$ basically influences the
overall ground-state phase diagram. In fact, the former case with
the antiferromagnetic Ising coupling $J_{0}=-1$ involves just four
different ground states: FM, SAF, AFM and FRU2. The phase boundary between 
FRU2 and FM(SAF) is delimited by the condition $|J_{1}|=1/2$, the phase boundary 
between FRU2 and AFM is given by $J_{z}=\sqrt{4J_{1}^{2}+J_{x}^{2}}$ and 
finally, the phase boundary between AFM and FM(SAF) is determined
by $J_{z}=J_{0}-2|J_{1}|+\sqrt{4J_{1}^{2}+J_{x}^{2}}$. Meanwhile, 
figure \ref{fig:The-phase-diagram}b illustrates the ground-state phase diagram 
for another particular case with the ferromagnetic Ising inter-leg coupling $J_{0}=1$, 
which displays the frustrated ground state 
FRU1 instead of the other frustrated ground state FRU2. The phase boundary
between FRU1 and AFM is delimited by $|J_{1}|=\frac{\sqrt{3}}{2}$, whereas
the phase boundary between FRU1 and FM (SAF) is given by $J_{z}=|J_{x}|-2|J_{1}|$,
and between AFM and FM(SAF) is $J_{z}=J_{0}-2|J_{1}|+\sqrt{4J_{1}^{2}+J_{x}^{2}}$.

\section{Thermodynamics}

To study the thermodynamics of the spin-1/2 Ising-Heisenberg ladder
with alternating inter-leg couplings, let us calculate first the partition
function given by 
\begin{equation}
\mathcal{Z}_{N}=\sum_{\{S\}}\left(\prod_{i=1}^{N}\mathrm{tr}_{i}\mathrm{e}^{-\beta\left(H_{i,i+1}^{XXZ}+H_{i,i+1}^{I}+H_{i,i+1}^{IH}\right)}\right).
\end{equation}
Here, $\beta=1/(k_{{\rm B}}T)$, $k_{{\rm B}}$ is being the Boltzmann's
constant, $T$ is the absolute temperature, the symbol $\mathrm{tr}_{i}$
denotes a trace over spin degrees of freedom of the $i$th Heisenberg
spin pair, the summation $\sum_{\{S\}}$ runs over all states of the
effective Ising spins $S=3/2$. After tracing out the spin degrees
of the Heisenberg spins one may employ the usual transfer-matrix approach
\cite{baxter-book} in order to calculate the partition function.
The transfer matrix $\boldsymbol{T}=\mathrm{tr}_{i}\Big(\mathrm{e}^{-\beta\left(H_{i,i+1}^{XXZ}+H_{i,i+1}^{I}+H_{i,i+1}^{IH}\right)}\Big)$
takes the following form 
\begin{equation}
\boldsymbol{T}=\left(\begin{array}{cccc}
w_{1,1} & w_{1,2} & w_{1,2} & w_{1,4}\\
w_{1,2} & w_{2,2} & w_{1,4}v^{-2} & w_{1,2}\\
w_{1,2} & w_{1,4}v^{-2} & w_{2,2} & w_{1,2}\\
w_{1,4} & w_{1,2} & w_{1,2} & w_{1,1}
\end{array}\right).\label{eq:T-trsnf}
\end{equation}
where individual matrix elements are explicitly given by 
\begin{alignat}{1}
w_{1,1}= & vz\left(y^{4}+y^{-4}\right)+\frac{v}{z}\left(x^{2}+x^{-2}\right),\\
w_{1,2}= & z\left(y^{2}+y^{-2}\right)+z^{-1}(y_{1}^{2}+y_{1}^{-2}),\\
w_{1,4}= & 2vz+\frac{v}{z}\left(x^{2}+x^{-2}\right),\\
w_{2,2}= & 2\frac{z}{v}+\frac{1}{vz}\left(y_{2}^{2}+y_{2}^{-2}\right),
\end{alignat}
with $x=\mathrm{e}^{\beta J_{x}/4}$ , $y=\mathrm{e}^{\beta J_{1}/4}$,
$z=\mathrm{e}^{\beta J_{z}/4}$, $v=\mathrm{e}^{\beta J_{0}/4}$,
$y_{1}=\mathrm{e}^{\beta\sqrt{J_{x}^{2}+J_{1}^{2}}/4}$, and $y_{2}=\mathrm{e}^{\beta\sqrt{J_{x}^{2}+4J_{1}^{2}}/4}$.

The eigenvalues $\lambda$ of the transfer matrix \eqref{eq:T-trsnf}
follow from the solution of eigenvalue problem $\text{det}(\boldsymbol{T}-\lambda)=0$.
The determinant drops into a fourth-order polynomial in $\lambda$,
which can be further factorized to 
\begin{alignat}{1}
\left(\lambda-w_{2,2}+\frac{w_{1,4}}{v^{2}}\right)\left(\lambda-w_{1,1}+w_{1,4}\right)\times\nonumber \\
\left(\lambda^{2}-2(p+q)\lambda+4pq-16r^{2}\right) & =0.\label{eq:cubic-eq}
\end{alignat}
The coefficients of quadratic polynomial are given by 
\begin{align}
p= & 2\left[{\rm e}^{\beta\frac{J_{0}+J_{z}}{4}}{\rm ch}\left(\tfrac{\beta J_{1}}{2}\right)^2+{\rm e}^{\beta\frac{J_{0}-J_{z}}{4}}{\rm ch}\left(\tfrac{\beta J_{x}}{2}\right)\right],\label{eq:cf-A}\\
q= & {\rm e}^{\frac{-\beta\left(J_{0}+J_{z}\right)}{4}}\left[{\rm ch}(\tfrac{\beta}{2}\sqrt{J_{x}^{2}+4J_{1}^{2}})+{\rm ch}(\tfrac{\beta J_{x}}{2})\right]+2{\rm e}^{\frac{-\beta\left(J_{0}-J_{z}\right)}{4}},\\
r= & {\rm e}^{\beta\frac{J_{z}}{4}}{\rm ch}\left(\tfrac{\beta J_{1}}{2}\right)+{\rm e}^{-\beta\frac{J_{z}}{4}}{\rm ch}\left(\tfrac{\beta}{2}\sqrt{J_{x}^{2}+J_{1}^{2}}\right).\label{eq:cf-C}
\end{align}
After that, one finds the following explicit form of the transfer-matrix
eigenvalues 
\begin{align}
\lambda_{0}= & p+q+\sqrt{(p-q)^{2}+16r^{2}},\label{eigmax}\\
\lambda_{1}= & p+q-\sqrt{(p-q)^{2}+16r^{2}},\\
\lambda_{2}= & \frac{\left(x^{2}y_{2}^{2}-1\right)\left(y_{2}^{2}-x^{2}\right)}{zx^{2}vy_{2}^{2}},\\
\lambda_{3}= & \frac{vz\left(y^{4}-1\right)^{2}}{y^{4}}.
\end{align}
It can be easily seen that the first eigenvalue (\ref{eigmax}) is
always positive and it always represents the largest eigenvalue of
the transfer matrix. In the thermodynamic limit $N\to\infty$, the
Helmholtz free energy per unit cell is given only by the largest transfer-matrix
eigenvalue through 
\begin{equation}
f=-\frac{1}{\beta}\ln\left(p+q+\sqrt{(p-q)^{2}+16r^{2}}\right),
\end{equation}
where $p$, $q$ and $r$ are given by Eqs. (\ref{eq:cf-A})-(\ref{eq:cf-C}).
The basic thermodynamic quantities as the entropy or specific heat
can be simply obtained from the Helmholtz free energy using the standard
thermodynamic relations.

\subsection{Entropy and specific heat}

\begin{figure}
\includegraphics[scale=0.43]{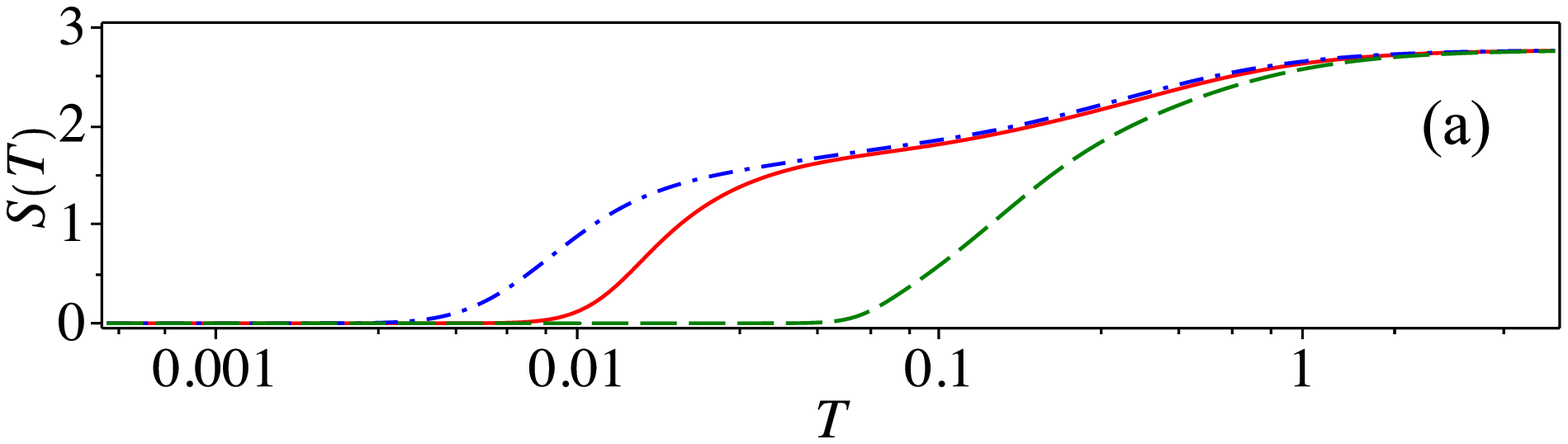} \includegraphics[scale=0.43]{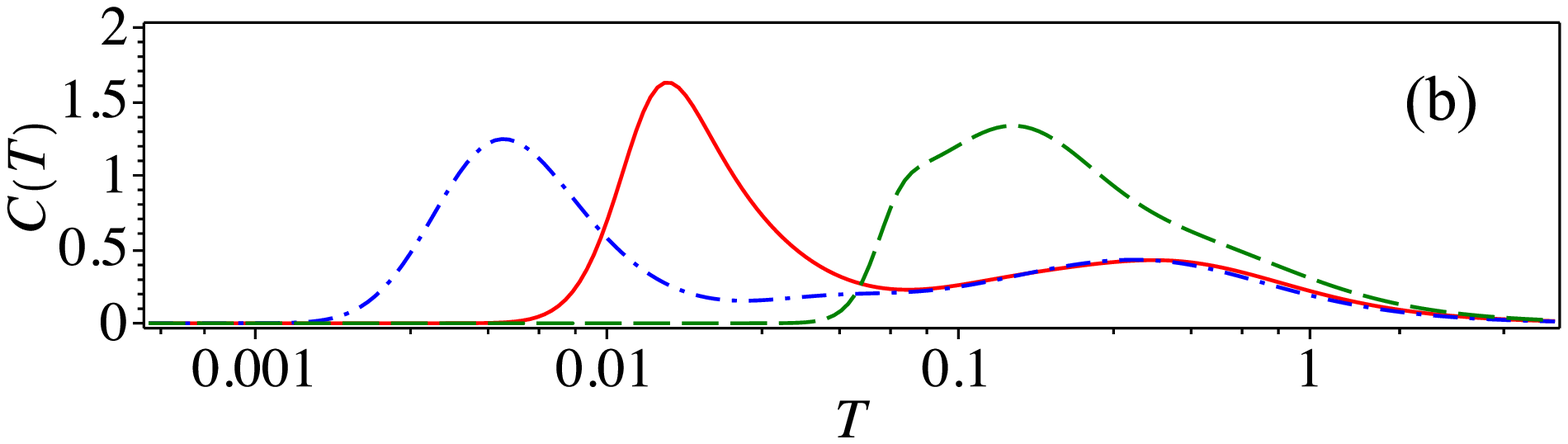}
\caption{\label{fig:entropy0}(a) The entropy as a function of the temperature
for the fixed values of the coupling constants $J_{0}=-1$ and $J_{x}=1$.
The solid line corresponds to a set of the interaction parameters
\{$J_{1}=0.56$, $J_{z}=1.4$\}, the dashed line corresponds to the
set \{ $J_{1}=0.4$, $J_{z}=1.23$\} and the dashed-dotted line corresponds
to the set \{$J_{1}=1.0$, $J_{z}=1.25$\}; (b) The corresponding
temperature dependence of the specific heat for the same set of the interaction parameters
in (a). }
\end{figure}

In figure \ref{fig:entropy0}(a) we illustrate temperature dependence
of the entropy for the fixed values of the coupling constants $J_{0}=-1$,
$J_{x}=1$. The choice of the interaction parameters \{$J_{1}=0.56$, $J_{z}=1.4$\}
drives the investigated model close to a triple coexistence point of the phases 
FM, AFM and FRU2, which lies in figure \ref{fig:The-phase-diagram}(a) 
at the coordinates \{$J_{1}=0.5$, $J_{z}=\sqrt{2}$\}. As one can
see, the temperature dependence of the entropy shows a steep increase
at low temperature $T\approx0.01$, which is followed by a gradual
temperature variation until another steeper change is reached at the
moderate temperature $T\approx0.2$. Similar behavior can be detected close to
the phase boundary of FRU2 and AFM when assuming fixed \{ $J_{1}=0.4$, $J_{z}=1.23$\},
as well as near the phase boundary of FRU2 and FM(SAF) assuming fixed \{$J_{1}=1.0$,
$J_{z}=1.25$\}.

These trends are also reflected in the corresponding thermal variations
of the specific heat, which are displayed in figure \ref{fig:entropy0}(b).
The specific heat evidently shows a pronounced double-peak temperature
dependence, whereas the low-temperature peak is relatively high and
sharp in a linear scale but it becomes round in a logarithmic scale.
Contrary to this, the high-temperature peak is broad both in a linear
as well as logarithmic scale. Obviously, the anomalous low-temperature
peak appears due to low-lying thermal excitations as all three phases
FM, AFM and FRU2 have equal energy at the triple point given by $J_{1}=0.5$
and $J_{z}=\sqrt{2}$.

\begin{figure}
\includegraphics[scale=0.43]{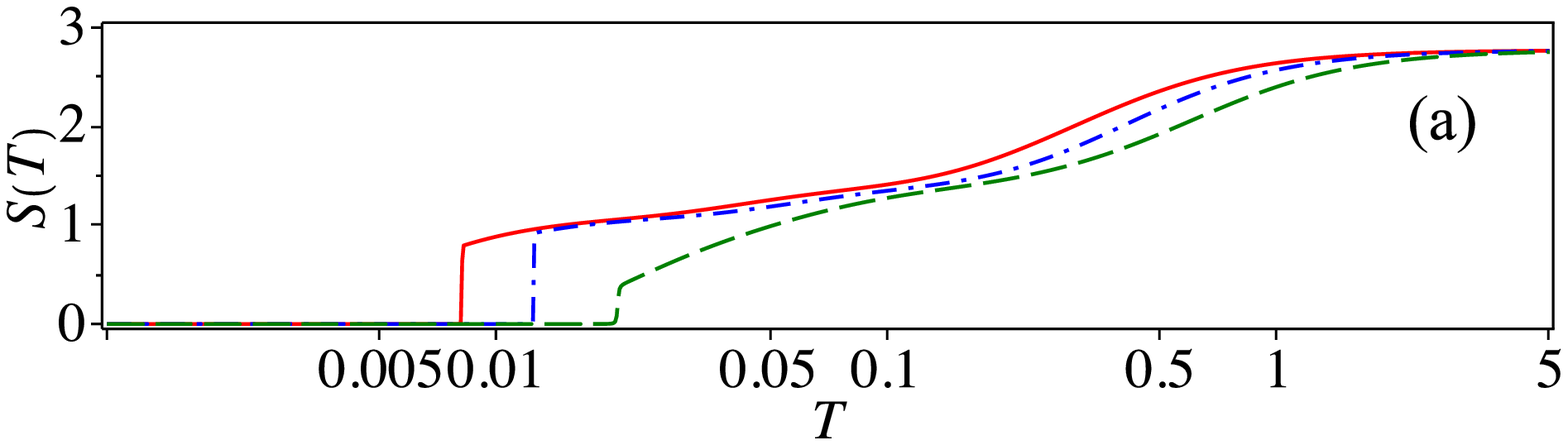} \includegraphics[scale=0.43]{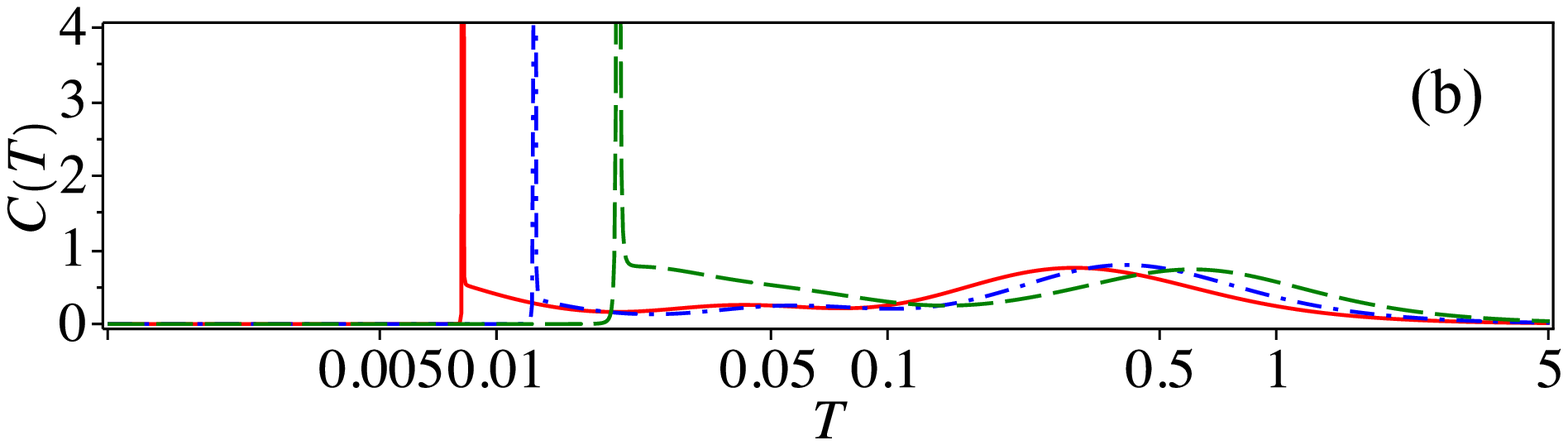}
\includegraphics[scale=0.43]{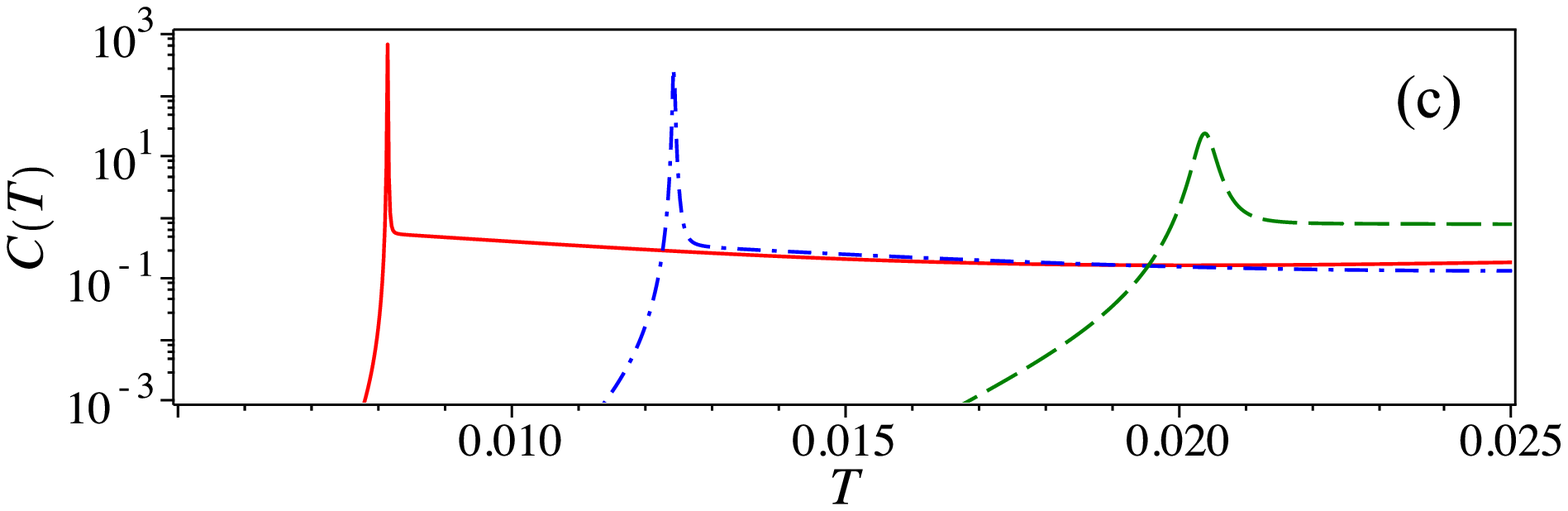}\caption{\label{fig:entropy1}(a) Thermal variations of the entropy for the
fixed value of the Ising inter-leg coupling $J_{0}=1$ and three different
sets of the interaction parameters. The solid line corresponds to
a set of the interaction parameters \{$J_{x}=0.5$, $J_{1}=0.724$,
$J_{z}=-0.9215$\}, the dashed line corresponds to the set \{$J_{x}=1$,
$J_{1}=0.89$, $J_{z}=-0.75$\} and the dashed-dotted line corresponds
to the set \{$J_{x}=2$, $J_{1}=1.2$, $J_{z}=-0.28$\}; (b) The specific
heat as a function of the temperature for the same set of parameters;
(c) The semi-logarithmic plot of the specific heat in a temperature
range, where a sharp low-temperature peak appears.}
\end{figure}

In figure \ref{fig:entropy1}(a) we display the entropy as a function
of temperature for the fixed value of the ferromagnetic Ising inter-leg coupling
$J_{0}=1$ and three different sets of the interaction parameters.
In all these plots one observes a unusual thermal behavior of the
entropy at sufficiently low temperature, where it shows an abrupt
but still continuous thermally-induced increase. This sudden increase
in the entropy is strongly reminiscent of the entropy jump, which
always accompanies a discontinuous (first-order) phase transition.
However, the abrupt but still continuous rise of the entropy appears
here owing to vigorous thermal excitations from two-fold degenerate
AFM ground state towards the macroscopically degenerate FRU1 state.
Therefore, the sudden rise of the entropy takes place at the temperature
\begin{equation}
T_{p}=\frac{\sqrt{4J_{1}^{2}+J_{x}^{2}}+J_{z}-|J_{x}|}{2\ln2},\label{tpeak}
\end{equation}
which can be obtained from a comparison of the Helmholtz free energy
of the AFM and FRU1 phases when simply ignoring a thermal change of
their internal energies. To provide a deeper insight, we have plotted
in figure \ref{fig:entropy1}(b) and (c) thermal variations of the
specific heat for the same set of parameters as for the entropy. The
specific heat exhibits remarkable double-peak temperature dependence
with a very sharp and narrow low-temperature maximum. The anomalous
specific-heat peak at low temperatures is strongly reminiscent because
of its extraordinary height and sharpness to an anomaly accompanying
a phase transition, but this peak is finite. The sharp low-temperature
peak of the specific heat can be thus identified with the Schottky-type
maximum \cite{gopal,karlova}, which is caused by intense thermal
excitations from the two-fold degenerate ground state AFM towards
the macroscopically degenerate excited state FRU1 driven by a high
entropy gain. As a matter of fact, the locus of the anomalous peak
is in accordance with the condition (\ref{tpeak}). 

\section{Bipartite entanglement}

Another fascinating topic that reserves its own right is a quantum
entanglement between the Heisenberg spin pairs. The quantity referred
to as the concurrence can be relatively simply adapted to quantify
the quantum entanglement between the spin-1/2 Heisenberg pair $\sigma_{a,i}$
and $\sigma_{b,i}$. The concurrence is defined through the reduced
density matrix $\rho$ \cite{wooters} 
\begin{equation}
\mathcal{C}(\rho)=\max\{{0,2\Lambda_{max}-\text{{tr}}\left(\sqrt{R}\right)}\},
\end{equation}
where 
\begin{equation}
R=\rho\sigma^{y}\otimes\sigma^{y}\rho^{*}\sigma^{y}\otimes\sigma^{y}.
\end{equation}
Above, $\Lambda_{max}$ is the largest eigenvalue of the matrix $\sqrt{R}$,
$\rho^{*}$ represent the complex conjugate of the reduced density
matrix $\rho$ and $\sigma^{y}$ is being the usual Pauli matrix.
The elements of the reduced density matrix \cite{bukman} can be expressed
in terms of the correlation functions \cite{amico}. Thus, the concurrence
is simply given by 
\begin{equation}
\mathcal{C}=\max\{0,4|\langle\sigma_{a}^{x}\sigma_{b}^{x}\rangle|-|\frac{1}{2}+2\langle\sigma_{a}^{z}\sigma_{b}^{z}\rangle|\}.\label{eq:Cnr}
\end{equation}
Two spatial components for the correlation function of the Heisenberg
spin pairs can be either obtained from the free energy, or equivalently
from the largest transfer-matrix eigenvalue using the following relation
\begin{alignat}{1}
\langle\sigma_{a}^{x}\sigma_{b}^{x}\rangle=\frac{1}{2\beta\lambda_{0}}\frac{\partial\lambda_{0}}{\partial J_{x}},\quad\text{and}\quad & \langle\sigma_{a}^{z}\sigma_{b}^{z}\rangle=\frac{1}{\beta\lambda_{0}}\frac{\partial\lambda_{0}}{\partial J_{z}}.
\end{alignat}
Alternatively, the concurrence for the Heisenberg spin pairs can be
written as 
\begin{equation}
\mathcal{C}=\frac{1}{2\beta\lambda_{0}}\max\{0,4|\frac{\partial\lambda_{0}}{\partial J_{x}}|-|\beta\lambda_{0}+4\frac{\partial\lambda_{0}}{\partial J_{z}}|\}.\label{eq:Cnr-rd}
\end{equation}

\subsection{Quantum entanglement}

\begin{figure}
\includegraphics[scale=0.4]{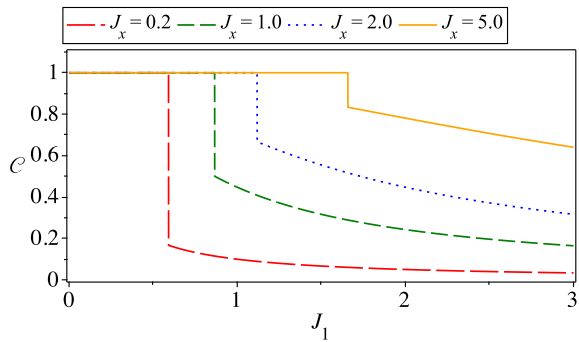}\caption{\label{fig:C-zero} Zero-temperature variations of the concurrence
with the Ising intra-leg interaction $J_{1}$ for several values of
the transverse component $J_{x}$ of the Heisenberg inter-leg interaction,
assuming fixed $J_{0}=-1$ and $J_{z}=-1$.}
\end{figure}

First, let us take a closer look at the ground-state behavior of the
concurrence. It is worthy to mention that the Heisenberg dimers are
maximally entangled ($\mathcal{C}=1$) at zero temperature just within
the FRU1 ground state. On the other hand, the AFM ground state also
shows at zero temperature the quantum entanglement of the Heisenberg
dimers when the concurrence depends on a relative strength of the
interaction parameters $J_{x}$ and $J_{1}$ 
\begin{equation}
\mathcal{C}=\frac{|J_{x}|}{\sqrt{4J_{1}^{2}+J_{x}^{2}}}.\label{eq:con-0}
\end{equation}
Contrary to this, the Heisenberg dimers are fully disentangled ($\mathcal{C}=0$)
within the other three classical ground states FM, SAF and FRU2. It
can be seen from figure \ref{fig:C-zero} that the zero-temperature
variations of the concurrence clearly demonstrate a first-order phase
transition from the FRU1 ground state to the AFM ground state through
the relevant discontinuity in the concurrence, assuming fixed $J_{0}=-1$
and $J_{z}=-1$. In general, the transverse component $J_{x}$ of
the Heisenberg inter-leg coupling enhances the concurrence, which
is contrarily suppressed by the Ising intra-leg interaction $J_{1}$.

\subsection{Thermal entanglement}

\begin{figure}
\includegraphics[scale=0.4]{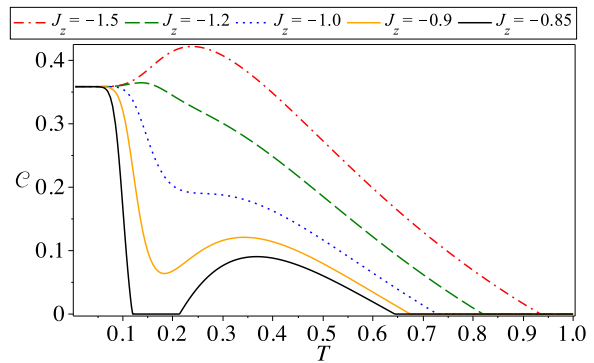}\caption{\label{fig:Conc-a} Temperature dependences of the concurrence for
the fixed values of the couplings constants $J_{0}=1$, $J_{x}=1$,
$J_{1}=1.3$ and several values of the interaction parameter $J_{z}$.}
\end{figure}

Next, let us discuss thermal entanglement of the Heisenberg dimers
at finite temperatures. In figure \ref{fig:Conc-a} we have plotted
the concurrence as a function of temperature for the set of parameters
driving the investigated system towards the AFM ground state. The
AFM ground state is entangled albeit not fully, because the concurrence
depends according to Eq. (\ref{eq:con-0}) on a competition between
the coupling constants $J_{x}$ and $J_{1}$. Figure \ref{fig:Conc-a}
illustrates an influence of the longitudinal component $J_{z}$ of
the Heisenberg inter-leg interaction on the concurrence at finite
temperature, which is however completely independent thereof at zero
temperature. In accordance with this statement, all displayed thermal
dependences of the concurrence tend towards the same zero-temperature
asymptotic limit given by Eq. \eqref{eq:con-0}. On the other hand,
it turns out that the concurrence is highly sensitive to the Heisenberg
coupling constant $J_{z}$ at higher temperature. Apart from a monotonous
decline of the concurrence with the rising temperature, one surprisingly
finds more peculiar non-monotonous thermal dependences of the concurrence
as shown in figure \ref{fig:Conc-a}.

\begin{figure}
\includegraphics[scale=0.4]{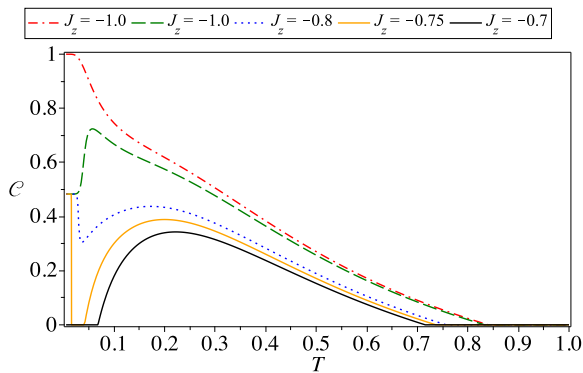}\caption{\label{fig:Conc-b}The concurrence as a function of temperature for
the fixed values of the interaction parameters $J_{0}=1$ and $J_{x}=1$.
The dashed-dotted (red) curve corresponds to a coexistence of the
AFM and FRU1 ground states at $J_{1}=\sqrt{3}/2$ and $J_{z}=-1$.
The other curves are plotted for $J_{1}=\sqrt{3}/2+0.04$ and different
values of $J_{z}$, which all fall into a parameter space of the AFM
ground state. }
\end{figure}

In figure \ref{fig:Conc-b}, we display one additional plot of the
concurrence exactly at and very close to a phase boundary between
the AFM and FRU1 ground states by keeping the coupling constants $J_{0}=1$
and $J_{x}=1$ fixed. The dashed-dotted (red) curve corresponds to
a coexistence of the AFM and FRU1 ground states, which occurs on assumption
that $J_{1}=\sqrt{3}/2$ and $J_{z}=-1$. As one can see, the concurrence
starts from its maximum asymptotic value $\mathcal{C}=1$ in this
particular case due to an infinite degeneracy of the FRU1 ground states.
The other temperature dependences of the concurrence are plotted in
figure \ref{fig:Conc-b} for $J_{1}=\sqrt{3}/2+0.04$ and different
values of $J_{z}$, which fall into a parameter space of the AFM ground
state. Owing to this fact, the zero-tempeture limit of the concurrence
dramatically falls to $\mathcal{C}\simeq0.4832$ in accordance with
Eq. \eqref{eq:con-0}. This sudden change is related to the zero-temperature
discontinuity of the concurrence at $J_{1}=\sqrt{3}/2$ provided that
the other three coupling constants $J_{0}=1$, $J_{1}=1$ and $J_{z}=-1$
are fixed.

The thermal entanglement within another parameter space, which corresponds
to the FRU1 ground state, exhibits standard temperature dependences
with a gradual monotonous temperature decline of the concurrence starting
from its maximum value $\mathcal{C}=1$ at zero temperature. From
this perspective, there is no need to display the standard thermal
variations of the concurrence within this parameter region.

\subsection{Threshold temperature}

The threshold temperature is one of the most relevant quantities used
for a characterization of the thermal entanglement, since it delimits
the entangled parameter space from the disentangled one. The threshold
temperature can be simply attained from Eq. \eqref{eq:Cnr-rd} when
letting the concurrence tend to zero from a non-zero side. Accordingly,
the threshold temperature can be obtained from a numerical solution
of the following transcendent (with respect to temperature) equation
\begin{equation}
4|\frac{\partial\lambda_{0}}{\partial J_{x}}|=|\beta\lambda_{0}+4\frac{\partial\lambda_{0}}{\partial J_{z}}|.
\end{equation}

\begin{figure}
\includegraphics[scale=0.4]{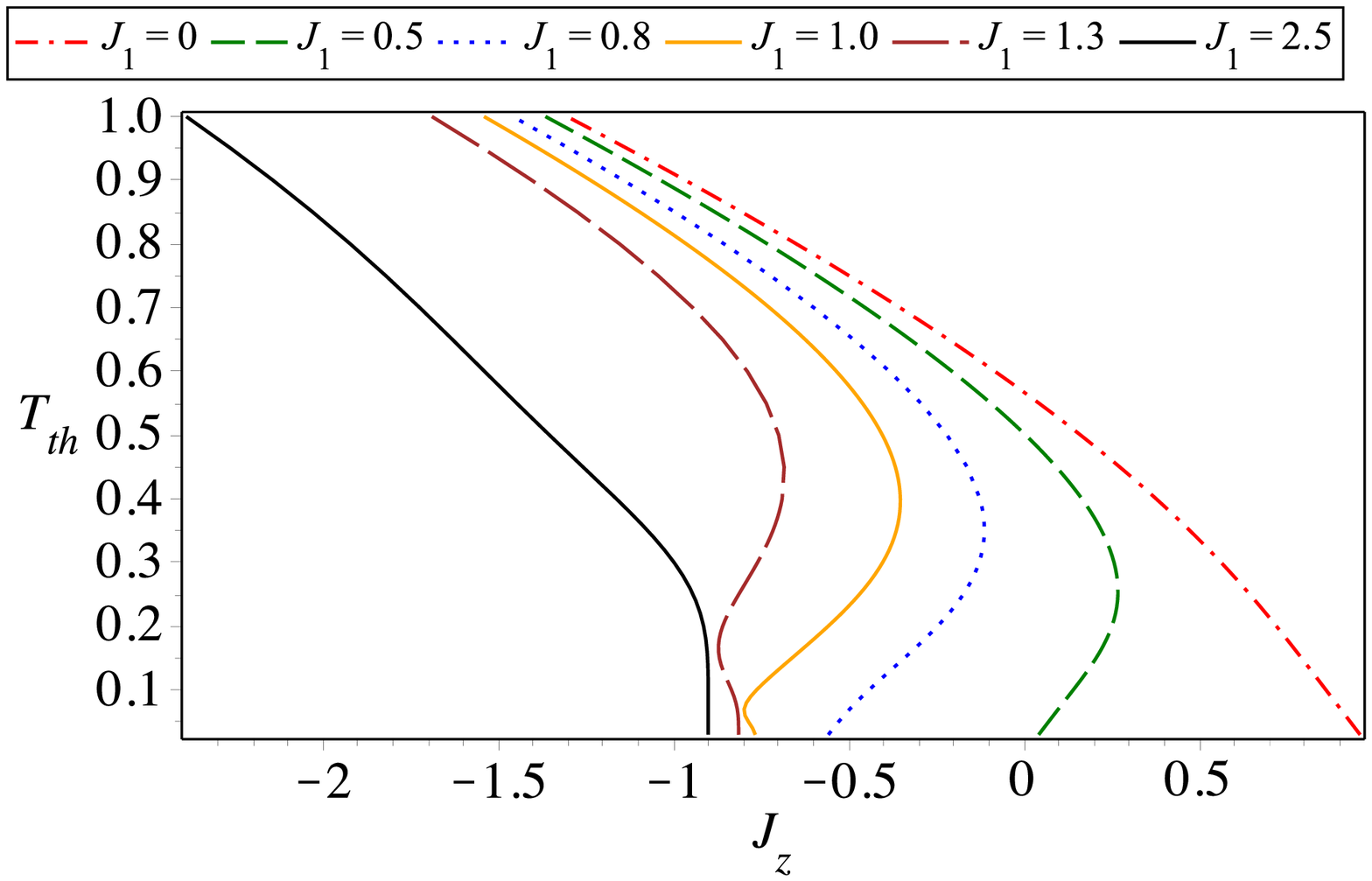}\caption{\label{fig:Thres-a}The threshold temperature $T_{th}$ against the
coupling constant $J_{z}$ for the fixed values of the interaction
parameters $J_{0}=1$, $J_{x}=1$ and several values of the Ising
intra-leg interaction $J_{1}$.}
\end{figure}

The threshold temperature $T_{th}$ is plotted in figure \ref{fig:Thres-a}
against the longitudinal component $J_{z}$ of the Heisenberg inter-leg
coupling by considering the fixed interaction parameters $J_{x}=1$,
$J_{0}=1$ and varying a strength of the Ising intra-leg coupling
$J_{1}$. The limiting case $J_{1}=0$ corresponds a set of non-interacting
Ising and Heisenberg dimers and hence, the threshold temperature exactly
coincides with that one of the spin-1/2 Heisenberg dimer that monotonically
decreases with $J_{z}$ until zero temperature is reached at the isotropic
Heisenberg point $J_{x}=J_{z}=1$. The relevant behavior of the threshold
temperature becomes much more complex for $J_{1}>0$, because it may
show a peculiar reentrant behavior when the entangled region re-appears
at temperatures above the disentangled region. The reentrant behavior
of the concurrence can be clearly seen for instance in figure \ref{fig:Conc-a}
for the parameter set $J_{0}=1$, $J_{x}=1$, $J_{z}=-0.85$ and $J_{1}=1.3$
(black line with three threshold temperatures). Apart from the triple
reentrance, the threshold temperature may also show double reentrance
(e.g. for $J_{1}=0.5$ in figure \ref{fig:Thres-a}) when the thermal
entanglement emerges above the disentangled ground state. It is worthy
to notice that the reentrant phenomenon disappear for the Ising intra-leg
couplings stronger than $J_{1}\gtrsim2.5$.

\begin{figure}
\includegraphics[scale=0.4]{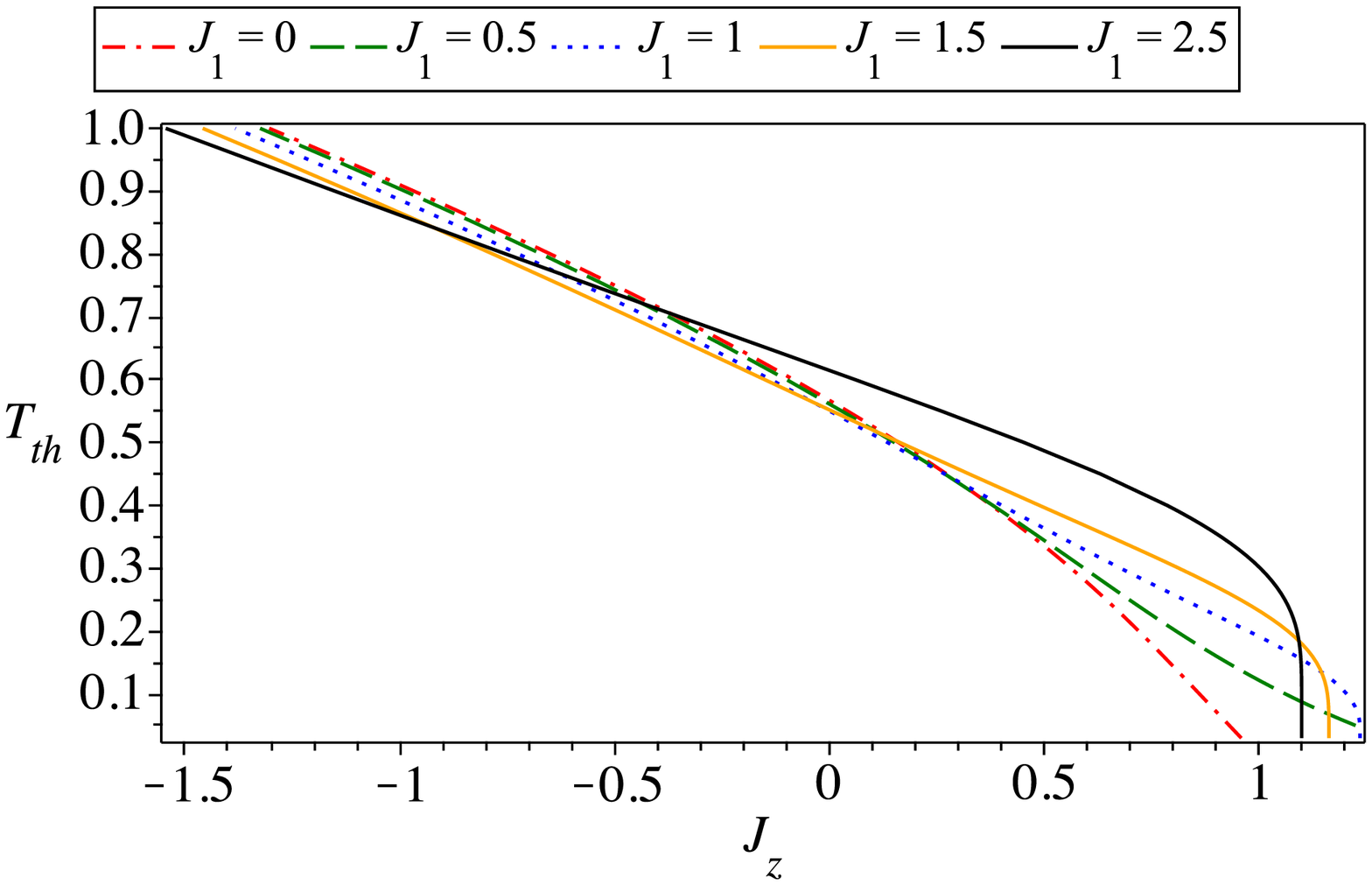}\caption{\label{fig:Thres-b}The threshold temperature $T_{th}$ against the
coupling constant $J_{z}$ for the fixed values of the interaction
parameters $J_{0}=-1$, $J_{x}=1$ and several values of the Ising
intra-leg interaction $J_{1}$.}
\end{figure}

Last but not least, the threshold temperature $T_{th}$ is displayed
in figure \ref{fig:Thres-b} as a function of the coupling constant
$J_{z}$ for the fixed values of the interaction parameters $J_{0}=-1$,
$J_{x}=1$ and several values of the Ising intra-leg coupling $J_{1}$.
The trivial case $J_{1}=0$ shown by the dashed-dotted (red) line,
which corresponds to the isolated spin-1/2 Heisenberg dimers, repeatedly
serve as a landmark to compare with. As soon as the Ising intra-leg
coupling $J_{1}$ is turned on, the threshold temperature reaches
zero at higher values of the coupling constant $J_{z}$, but afterwards
it recovers the zero-temperature asymptotic value $J_{z}\rightarrow1$
for strong enough Ising intra-leg couplings $J_{1}\gg1$.

\section{Conclusion}

In the present work, we have exactly solved the spin-1/2 Ising-Heisenberg
ladder accounting for regularly alternating Ising and Heisenberg inter-leg
couplings in addition to the Ising intra-leg interaction. It has been
evidenced that the investigated model is equivalent to the mixed spin-(3/2,1/2)
Ising-Heisenberg diamond chain with the nodal Ising spin $S=3/2$
and the interstitial spin-1/2 Heisenberg dimers, which was exactly
treated by means of the transfer-matrix method. Using this rigorous
procedure, we have found that the ground-state phase diagram involves
in total five different ground states: ferromagnetic, antiferromagnetic,
super-antiferromagnetic and two types of highly degenerated (frustrated)
ground-state manifolds. The antiferromagnetic and one of frustrated
ground states are quantum in character as exemplified by the quantum
and thermal entanglement of the Heisenberg dimers. In addition, the
concurrence as a measure of the thermal entanglement may exhibit a
striking reentrant behavior.

We have also exactly calculated the entropy and specific heat, which
may display under certain conditions anomalous thermal dependences.
The entropy may exhibit at sufficiently low temperatures an abrupt
but still continuous rise, which gives rise to an extraordinary high
and sharp specific-heat maximum. The relevant temperature dependences
of the entropy and specific heat thus mimic in many respects a temperature-driven
phase transition, but they should not be confused as signatures of
it. The anomalous thermal behavior of the entropy and specific heat
occurs in the present model due to a high entropy gain, which originates
from vigorous thermal excitations between the two-fold degenerate
ground state and the highly degenerate excited state close enough
in energy. The model under investigated thus falls into a prominent
class of the exactly solved systems with such an intriguing magnetic
behavior \cite{gali,alec}.

\end{document}